\documentclass[aps,reprint,superscriptaddress]{revtex4-2}
\usepackage{newtxtext,newtxmath}
\usepackage{wrapfig}
\usepackage{bm}
\usepackage{graphicx} 
\usepackage[normalem]{ulem}
\usepackage{xcolor}

\newcommand{\Add}[1]{}%{\textcolor{magenta}{#1}}
\newcommand{\Del}[1]{}%{\textcolor{blue}{\sout{#1}}}
\newcommand{\Note}[1]{}%{\textcolor{red}{\bf #1}}
\begin{document}

\title{Dynamical Theory of Elastic Synchronization of Cardiomyocytes}

%%\author{Akinari Tomiie and Nariya Uchida\thanks{nariya.uchida@tohoku.jp}}
%%\inst{Department of Physics, Tohoku University, Sendai 980-8578, Japan } %\\

\author{Akinari Tomiie}
\affiliation{Department of Physics, Tohoku University, Sendai 980-8578, Japan}

\author{Nariya Uchida}
\affiliation{Department of Physics, Tohoku University, Sendai 980-8578, Japan}
\email{nariya.uchida@tohoku.jp}

%%\abst{
\begin{abstract}
We study synchronization of two cardiomyocytes mediated by elastic interactions through the substrate. Modeling each cell as an oscillating force dipole governed by a Rayleigh-type equation, we derive an effective mechanical coupling from the elastic response of the surrounding medium. Using phase reduction theory, supported by direct numerical simulations, we obtain a dynamical phase description for two cardiomyocytes that predicts geometry-dependent selection of synchronized states. Depending on the mutual orientation, the cells robustly converge to either in-phase or anti-phase beating, yielding an orientation-dependent state map with a nontrivial state boundary. The synchronization time also depends strongly on the distance and mutual orientation of the cells. These results bridge earlier energetic two-body theory and dynamical single-cell theory, and provide a dynamical framework for elastic synchronization of cardiomyocytes.
\end{abstract}
%%}

\maketitle
%%%%%%%%%%%%%%%%%%%%%%%%%%%%%%%%%%%%%%%%%%%%%%%%%%%%%%%%%%%%%%%%%%
%\section{Introduction}

The synchronized beating of cardiomyocytes is essential for efficient
cardiac function. In intact cardiac tissue, such coordination is usually
attributed to electrochemical signaling mediated by pacemaker cells and
intercellular coupling~\cite{bers2002, haraguchi2006}.
However, recent experiments have shown that even isolated cardiomyocytes
can synchronize through mechanical deformations of the surrounding elastic
substrate, without direct electrical or chemical communication~\cite{tang2011, nitsan2016}.
Such cells can also be mechanically paced by an external oscillatory
probe, with the induced response developing over a timescale of order
10--15 min~\cite{nitsan2016}.
Furthermore, self-beating cardiac spheroids have been found to synchronize
through mechanical communication alone~\cite{nakano2021mechanical}.
These observations indicate that mechanical interactions can play an
important role in cardiac synchronization and motivate a theoretical
description of beating cells as mechanically coupled active oscillators.

Previous theoretical studies have provided two important foundations for this problem. First, Cohen and Safran modeled cardiomyocytes as oscillating force dipoles embedded in an elastic or viscoelastic medium, and showed that the preferred synchronized state depends on the mutual orientation of the cells~\cite{cohen2016}. Their analysis demonstrated that elastic interactions can favor either in-phase or anti-phase beating depending on the geometry, thereby clarifying the energetic basis of mechanically selected synchronized states. However, the main focus of that work was the steady-state selection of synchronized phases rather than the explicit time evolution toward synchronization.

Second, Cohen and Safran later developed a nonlinear oscillator theory for a mechanically driven cardiomyocyte based on a Rayleigh-type equation~\cite{cohen2018}. Using an Adler-type phase description, they analyzed entrainment by an oscillating external mechanical probe and explained spontaneous, entrained, and bursting responses of a single paced cell. This framework established the usefulness of a coarse-grained nonlinear oscillator description for mechanically driven cardiomyocytes, but it was formulated primarily for probe-driven single-cell dynamics and did not explicitly address configuration-dependent coupling between two interacting cells. The use of such a nonlinear active-oscillator description is also motivated by earlier theoretical studies of spontaneous cellular oscillations and acto-myosin dynamics~\cite{julicher1997,julicher2001}.
A related theoretical study derived a van der Pol-type effective equation
for spontaneous calcium oscillations in cardiac cells and showed that it can
account for entrainment by external mechanical or electrical forcing~\cite{cohen2019physics}.

These studies motivate a dynamical theory that combines nonlinear limit-cycle oscillations of individual cardiomyocytes with orientation-dependent elastic interactions between a pair of cells. Developing such a theory is important for connecting the energetic picture of mechanically selected synchronized states with the actual phase dynamics of mechanically coupled cardiomyocytes.

Rather than attempting a detailed molecular description of acto-myosin
activity and calcium dynamics, we adopt a phenomenological nonlinear-oscillator
framework for cardiomyocyte beating~\cite{julicher1997,julicher2001}.
In this study, we formulate such a dynamical model by combining a
Rayleigh-type description of cardiomyocyte oscillations~\cite{cohen2018}
with geometry-dependent elastic coupling mediated by the
substrate~\cite{cohen2016}.
We then analyze the resulting synchronization dynamics using both direct
numerical simulations and phase reduction theory.

%%%%%%%%%%%%%%%%%%%%%%%%%%%%%%%%%%%%%%%%%%%%%%%%%%%%%%%%%%%%%%%%%%%%%%%%
%\section{Model}
%\subsection{Rayleigh-type dynamics of a cardiomyocyte}

To describe the spontaneous beating of a cardiomyocyte, we introduce a
coarse-grained variable $X(t)$ representing the extension of the cell
along its long axis.
We assume that $X(t)$ obeys a Rayleigh-type equation
\begin{equation}
\gamma \ddot{X} + \alpha \dot{X} + \beta \dot{X}^{3} + kX = 0.
\label{eq:rayleigh_dim}
\end{equation}
\Add{Here, the coefficient $\gamma$ plays the role of an effective mass, 
or phenomenological inertial coefficient, representing the delayed response of the coarse-grained contractile dynamics rather than the physical mass of the cell. The parameter $k$ is an effective stiffness, and $\alpha$ and $\beta$ characterize the linear and nonlinear velocity-dependent terms, respectively.}
\Del{Here, the coefficient $\gamma$ plays the role of an effective mass,
$k$ is an effective stiffness, and $\alpha$ and $\beta$ characterize
the linear and nonlinear velocity-dependent terms, respectively.}
For $\alpha<0$ and $\beta>0$, the linear term provides self-excitation at small amplitude, whereas the cubic term suppresses growth at large amplitude, yielding a stable self-sustained oscillation. This coarse-grained use of a Rayleigh-type equation follows the same spirit as earlier nonlinear descriptions of mechanically driven cardiomyocytes~\cite{cohen2018}.

Introducing a characteristic length scale $L_0$ and frequency
$\omega=\sqrt{k/\gamma}$, we define the dimensionless variables
\begin{equation}
\hat{X}=\frac{X}{L_{0}}, \qquad \tau=\omega t.
\end{equation}
Equation~(\ref{eq:rayleigh_dim}) is then rewritten as
\begin{equation}
\frac{d^{2}\hat{X}}{d\tau^{2}}
-\mu \left[ 1-\left(\frac{d\hat{X}}{d\tau}\right)^{2} \right] 
\frac{d\hat{X}}{d\tau} +\hat{X} = 0,
\label{eq:rayleigh_nondim}
\end{equation}
where $\mu=-\alpha/\sqrt{\gamma k}$. 
Eq.~(\ref{eq:rayleigh_nondim}) has a stable limit cycle, representing an isolated cardiomyocyte in our model.

%%%%%%%%%%%%%%%%%%%%%%%%%%%%%%%%%%%%%%%%%%%%%%%%%%%%%%%%%%%%%
%\subsection{Force-dipole representation and elastic coupling}

Next, we consider two cardiomyocytes adhered to the surface of a semi-infinite elastic substrate occupying the half-space below the surface. The cells are assumed to lie in the surface plane, so that their spatial arrangement is described in two dimensions. 
We choose the $x$-axis along the line connecting the two cells, so
that the configuration is specified by the center-to-center
distance $d$ and the orientation angles $\theta_1$ and $\theta_2$
relative to the $x$-axis.
The geometry of the model is illustrated schematically in Fig.~1.

%%%%%%%%%%%%%%%%%%%%
\begin{figure}[htbp]
  \centering
  \includegraphics[width=0.8\linewidth]{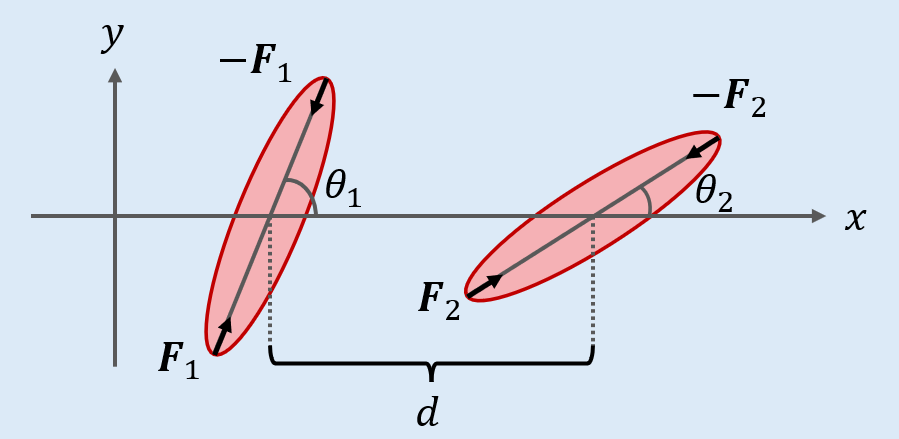}
  \caption{Geometry of a pair of cardiomyocytes modeled as force dipoles on an elastic substrate. The configuration is specified by the center-to-center distance $d$ and the orientation angles $\theta_1$ and $\theta_2$.}
  \label{fig:sample2}
\end{figure}
%%%%%%%%%%%%%%%%%%%%

Each cardiomyocyte is represented as a force dipole acting along its long axis. For cell $i=1,2$, let $\bm{e}_i=(\cos\theta_i,\sin\theta_i)$ be the unit vector specifying its orientation and $\bm{r}_{i,c}$ its center position
with $\bm{r}_{2,c} - \bm{r}_{1,c} = (d,0)$.
The extension $X_i(t)$ is incorporated into the dipole geometry through the pole positions 
\begin{equation}
\bm{r}_i^{\pm}(t)=\bm{r}_{i,c}
\pm \left(\frac{l_{\rm cell}+X_i(t)}{2}\right)\bm{e}_i ,
\end{equation}
where $l_{\rm cell}$ is the resting cell length.

The force density generated by cell $i$ is modeled as
\begin{equation}
\bm{f}_i(\bm{r},t)
=
F_i(t)
\left[
S(\bm{r}-\bm{r}_i^{+}(t)) - S(\bm{r}-\bm{r}_i^{-}(t))
\right]
\bm{e}_i,
\end{equation}
where \Add{$F_i(t)$ is the force exerted by each pole on the substrate, and} 
$S(\bm{r})$ is a {normalized} Gaussian shape function defined by
\begin{equation}
S(\bm{r})
=
{\frac{1}{2\pi \sigma_{\rm cell}^2}}
\exp\!\left(
-\frac{|\bm{r}|^2}{2\sigma_{\rm cell}^2}
\right),
\end{equation}
with $\sigma_{\rm cell}$ setting the spatial width of the force profile.

The substrate is modeled as a semi-infinite linear elastic medium characterized by Young's modulus $E$ and Poisson's ratio $\nu$. Since the cells are attached to its surface, we consider the in-plane surface displacement induced by 
the total force density $\bm{f}^{\rm (tot)}(\bm{r},t) = \bm{f}_1(\bm{r},t) + \bm{f}_2(\bm{r},t)$,  
\begin{equation}
u_{\alpha}(\bm{r},t)
=
\sum_{\beta} \int G_{\alpha\beta}(\bm{r}-\bm{r}')
\,f^{\rm (tot)}_{\beta}(\bm{r}',t)
\,d\bm{r}' ,
\end{equation}
where $\alpha,\beta = x,y$ and
$G_{\alpha\beta}(\bm{r})$ is the corresponding surface Green tensor,~\cite{landau1986}
\begin{equation}
G_{\alpha\beta}(\bm{r})
=
\frac{1+\nu}{2\pi E}
\left[
\frac{2(1-\nu)}{r}\delta_{\alpha\beta}
+
\frac{2\nu r_{\alpha}r_{\beta}}{r^3}
\right].
\end{equation}
\Add{Because the cell extensions $X_i$ and the dipole forces $F_i$ are not independent, 
the coupled dynamics effectively has two degrees of freedom, $X_1$ and $X_2$.}
To determine the feedback from the substrate to the cells, we introduce a compliance matrix $\bm{M}$ relating the dipole amplitudes to the cell extensions,
\begin{equation}
X_i=\sum_j M_{ij}F_j .
\label{eq:compliance_relation}
\end{equation}
Its elements are obtained from the projected relative displacement of the two poles of cell $i$ induced by a unit dipole force applied to cell $j$,
\begin{equation}
M_{ij}={w_{ij}^{+}-w_{ij}^{-}}, \qquad
w_{ij}^{\pm}=\bm{e}_i\cdot \bm{u}^{(j)}(\bm{r}_i^{\pm},t).
\end{equation}
Here $\bm{u}^{(j)}$ denotes the displacement field induced by cell $j$. 
Inverting Eq.~(\ref{eq:compliance_relation}), we obtain the forces 
as functions of the cell extensions,
\begin{equation}
\bm{F}=\bm{M}^{-1}\bm{X},
\label{eq:F_from_X}
\end{equation}
where $\bm{X}=(X_1,X_2)^{\mathrm T}$ and $\bm{F}=(F_1,F_2)^{\mathrm T}$.
The force acting on each cell is the reaction force from the substrate.
The dynamics of two elastically coupled cardiomyocytes is then written as
\begin{equation}
\frac{d^{2}\hat{X}_i}{d\tau^{2}}
-\mu
\left[
1-\left(\frac{d\hat{X}_i}{d\tau}\right)^{2}
\right]
\frac{d\hat{X}_i}{d\tau}
+\hat{X}_i
=
-\hat{F}_{i}(\hat{X}_1,\hat{X}_2),
\label{eq:coupled_rayleigh}
\end{equation}
where $\hat{F}_i = F_i/(kL_0)$.
The negative sign reflects the fact that the force exerted on the
cell is opposite to that exerted by the cell on the substrate.
Note that $\hat{F}_i$ contains both self-response and mutual-interaction contributions.
Since the self-response terms diverge in the point-force limit ($\sigma_{\rm cell}\to 0$), 
we retain a finite width $\sigma_{\rm cell}$ comparable to the cell width.

%%%%%%%%%%%%%%%%%%%%%%%%%%%%%%%%%%%%%%%%%%%%%%%%%%%%%%%%%%%%%
%\subsection{Phase reduction theory}

We consider a parameter regime in which the substrate-mediated interaction is weak, so that the phase-coupling term satisfies $|\Gamma_i(\Delta\phi)| \ll \Omega$
and the phase dynamics provides a closed description of synchronization.
For an uncoupled Rayleigh oscillator, the periodic solution
$\hat{X}_0(\tau)$ defines a stable limit cycle in the \Add{state} \Del{phase} space spanned by
$\bm{q}=(\hat{X}, d\hat{X}/d\tau)^{\mathrm T}$.
A phase variable $\phi$ is introduced along this limit cycle so
that $\phi$ increases by $2\pi$ in a single period $T_0$ and
$d\phi/d\tau=\Omega_0$ for the unperturbed dynamics, where
$\Omega_0=2\pi/T_0$ is the constant intrinsic frequency.

For weak coupling, we treat the elastic reaction force in
Eq.~\eqref{eq:coupled_rayleigh} as a perturbation to this
limit-cycle dynamics. 
\Add{We define the state vector and perturbation vector as}
\Del{Using the state vector and perturbation vector}
\begin{equation}
\bm{q}_i=
\left(
\hat{X}_i,
\frac{d\hat{X}_i}{d\tau}
\right)^{\mathrm T},
\qquad
\bm{P}_i=
\begin{pmatrix}
0\\
-\hat{F}_i
\end{pmatrix}.
\end{equation}
\Add{The perturbation vector $\bm{P}_i$ expresses the elastic-interaction term 
in the first-order dynamical system obtained by rewriting 
Eq.~\eqref{eq:coupled_rayleigh}.}
\Add{For} \Del{for} cell $i$, the phase dynamics is written as
\begin{equation}
\frac{d\phi_i}{d\tau}
=
\Omega_0+\bm{Z}(\phi_i)\cdot
\bm{P}_i\!\left(\hat{X}_0(\phi_1),\hat{X}_0(\phi_2)\right).
\label{eq:phase_eq_preavg}
\end{equation}
Here, $\bm{Z}(\phi)$ is the phase sensitivity function~\cite{kuramoto1984},
\Add{computed from the adjoint equation for the unperturbed limit cycle 
$\bm q_0(\phi)=(\hat X_0(\phi),d\hat X_0/d\tau)^{\mathrm T}$, 
with periodic boundary conditions and the normalization 
$\bm Z(\phi)\cdot d\bm q_0/d\tau=\Omega_0$.}
\Del{and the} \Add{The} dot denotes the inner product between the two-component vectors 
$\bm Z$ and $\bm P_i$.
The perturbation term $\bm{Z}(\phi_i)\cdot\bm{P}_i$ consists of a local
self-response part and a mutual-interaction part. The former is absorbed
into an effective frequency $\Omega$, while the latter determines the
phase coupling between the two cells. 
We define the interaction part of the perturbation vector 
\begin{equation}
\bm{P}_i^{\rm (int)}=
\begin{pmatrix}
0\\
-\hat{F}_i^{\rm (int)}
\end{pmatrix},
\end{equation}
where $\hat F_i^{\rm (int)}$ denotes the interaction part of the reaction force on cell $i$, induced by the other cell.

To obtain a reduced equation for the phase difference
$\Delta\phi=\phi_2-\phi_1$, we write $\phi_1=\psi$ and
$\phi_2=\psi+\Delta\phi$, and average over one period of
the fast phase $\psi$ while keeping $\Delta\phi$ fixed.
This yields the averaged phase dynamics
\begin{equation}
\frac{d\phi_i}{d\tau} = \Omega + \Gamma_i(\Delta\phi),
\end{equation}
with
\begin{flalign}
\Gamma_1(\Delta\phi)
&=
\frac{1}{2\pi}
\int_0^{2\pi} \!
\bm{Z}(\psi)\cdot
\bm{P}_1^{\rm (int)}\!\left(\hat{X}_0(\psi),\hat{X}_+(\psi)\right)\,d\psi ,
\\
\Gamma_2(\Delta\phi)
&=
\frac{1}{2\pi}
\int_0^{2\pi} \!
\bm{Z}_+(\psi)\cdot
\bm{P}_2^{\rm (int)}\!\left(\hat{X}_0(\psi),\hat{X}_+(\psi)\right)\,d\psi,
\end{flalign}
where we used $\bm{Z}_+(\psi)=\bm{Z}(\psi+\Delta\phi)$ and
$\hat{X}_+(\psi)=\hat{X}_0(\psi+\Delta\phi)$ for brevity.
The phase difference then obeys
\begin{equation}
\frac{d(\Delta\phi)}{d\tau}
=
\Gamma_2(\Delta\phi)-\Gamma_1(\Delta\phi)
\equiv
\Gamma(\Delta\phi).
\end{equation}
Stable phase locking occurs at fixed points $\Delta\phi^\ast$ satisfying
\begin{equation}
\Gamma(\Delta\phi^\ast)=0,
\qquad
\left.
\frac{d\Gamma}{d(\Delta\phi)}
\right|_{\Delta\phi=\Delta\phi^\ast}
<0.
\label{eq:stability}
\end{equation}
This reduced description allows us to predict the selected synchronized state from the geometry-dependent elastic coupling.

%%%%%%%%%%%%%%%%%%%%%%%%%%%%%%%%%%%%%%%%%%%%%%%%%%%%%%%%
%\section{Results}
%\subsection{Parameters}

Next, we present the synchronization dynamics obtained from phase
reduction theory and direct numerical simulations. Experimentally,
substrate-mediated synchronization has been reported for substrates
with Young's modulus of order $1~{\rm kPa}$,~\cite{tang2011,nitsan2016,engler2008}
comparable to the stiffness of neonatal cardiomyocytes or the surrounding
extracellular matrix (ECM)~\cite{jacot2010,bhana2010}.
The effective stiffness $k$ is typically of order
$0.1$--$500~{\rm nN}/\mu{\rm m}$~\cite{taylor2013,rodriguez2014,ribeiro2015}.
In our simulations, we 
{introduce $k' = k/(2\pi \sigma_{\rm cell}^2)$ and} 
set {$k'=1~{\rm nN}/\mu{\rm m}^3$,}
$l_{\rm cell}=20~\mu{\rm m}$,
$\sigma_{\rm cell}=5~\mu{\rm m}$, $E=1~{\rm kPa}$, $\nu=0.49$, and $\mu=1$.
{This gives the stiffness $k \approx 157 \, \mathrm{nN}/\mu \mathrm{m}$, 
which is within the range of experimental values.}
With $L_0=1~\mu{\rm m}$ and $\omega=1~{\rm s}^{-1}$, the corresponding
dimensionless {length scales} are  $\hat l_{\rm cell}=20$ 
and $\hat \sigma_{\rm cell}=5$, and the distance is varied in the range
$40 \le \hat d \le 120$. 
For the present parameter range, the nonlinearity and the self-response
modify the oscillation period only weakly, with $\omega T_0/(2\pi)\approx 1.06$ and $T/T_0\approx 0.97$.

%\subsection{Comparison between direct simulation and phase reduction results}

%%%%%%%%%%%%%%%%%%%%%%
\begin{figure}[bp]
  \centering
  \includegraphics[width=1.0\linewidth]{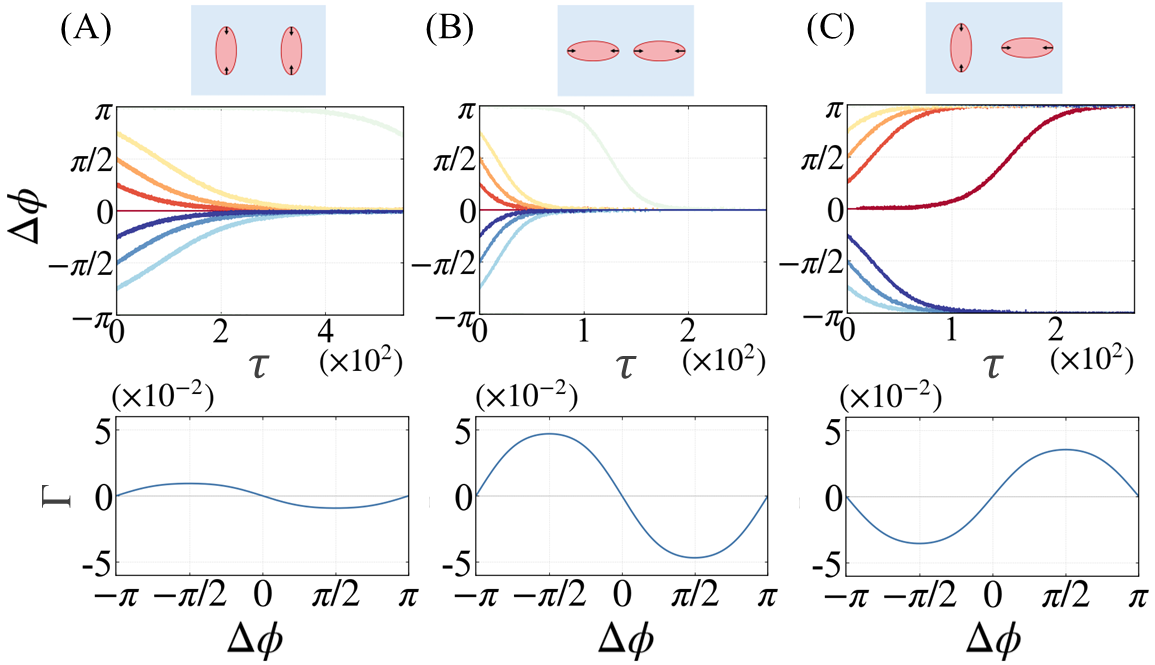}
  \caption{Synchronization patterns for the representative configurations (A), (B), and (C). The top row shows the phase difference
$\Delta\phi=\phi_2-\phi_1$ from direct numerical simulations, and the
bottom row shows the corresponding phase-coupling function
$\Gamma(\Delta\phi)$ from phase reduction theory. The distance is set to
$d/l_{\rm cell}=2$ for (A) and (C), and $d/l_{\rm cell}=2.5$ for (B).
\Add{In (C), the red trajectory shows a slow relaxation 
toward the anti-phase synchronized state because the initial phase difference 
is close to the unstable fixed point $\Delta \phi=0$.}
}
\label{fig:comparison}
\end{figure}
%%%%%%%%%%%%%%%%%%%%%%

We first consider three representative configurations:
(A) $\theta_1=\theta_2=\pi/2$ with $d/l_{\rm cell}=2$,
(B) $\theta_1=\theta_2=0$ with $d/l_{\rm cell}=2.5$, and
(C) $\theta_1=\pi/2$, $\theta_2=0$ with $d/l_{\rm cell}=2$.
We performed direct numerical simulations of the coupled dynamics [Eqs.~\eqref{eq:F_from_X} and \eqref{eq:coupled_rayleigh}] using the fourth-order Runge-Kutta method with a time step of $\Delta t = 0.1$. We also calculated the corresponding phase coupling function $\Gamma(\Delta\phi)$ based on phase reduction theory. 
By tracking the relative phase difference, $\Delta\phi=\phi_2-\phi_1$, both the direct numerical simulations and the phase reduction analysis show that configurations (A) and (B) converge to states corresponding to in-phase ($|\Delta\phi|=0$) synchronization, while (C) converges to a state corresponding to anti-phase ($|\Delta\phi|=\pi$) synchronization (Fig.~\ref{fig:comparison}). These results reproduce the predictions of previous energetic analyses, which suggested that the substrate's elastic energy is minimized for in-phase oscillation in (A) and (B), and for anti-phase oscillation in (C)~\cite{cohen2016}. The agreement between direct numerical simulations and phase reduction theory supports the validity of the reduced description in the present weak-coupling regime.
For these cases, the phase-coupling function is nearly sinusoidal, and thus
the characteristic synchronization time scales as $\tau_{\rm sync}\sim 1/\Delta \Gamma$,
where $\Delta \Gamma \equiv \Gamma_{\max}-\Gamma_{\min}$ is the peak-to-peak amplitude
of the coupling function. This explains the slower synchronization in configuration
(A) than in (B) and (C) in the direct numerical simulations, a feature not predictable
from the energy-minimization principle.

%%%%%%%%%%%%%%%%%%%%%%
\begin{figure}[htbp]
  \centering
  \includegraphics[width=0.8\linewidth]{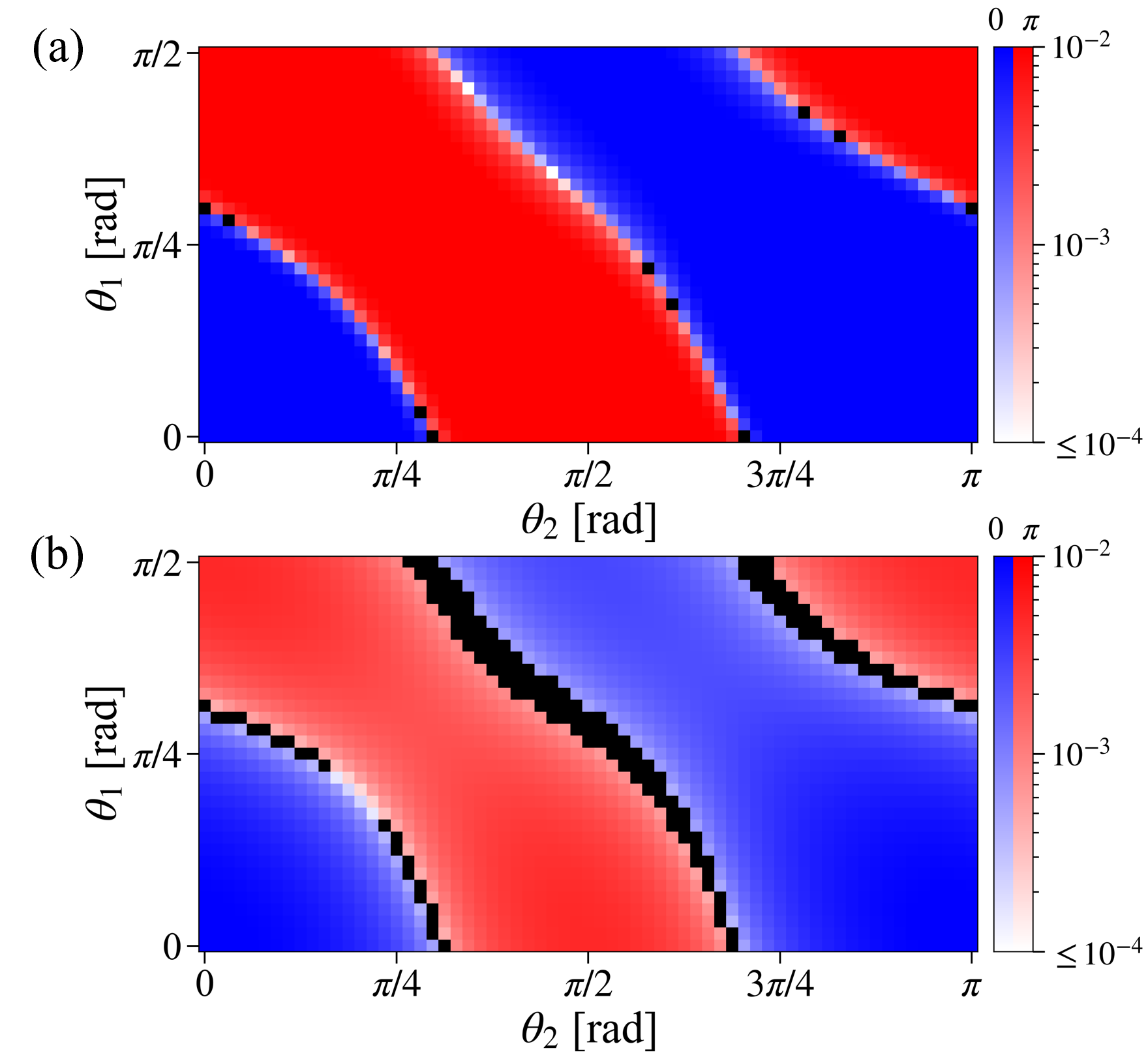}
  \caption{Orientation-dependent dynamical state maps for (a) $d/l_{\rm cell}=2.5$ and (b) $d/l_{\rm cell}=6$. Blue and red regions represent configurations yielding in-phase ($|\Delta \phi^*| < 0.1$) and anti-phase ($|\pi - |\Delta \phi^*|| < 0.1$) synchronization, respectively. In these regions, the brightness indicates the coupling amplitude $\Delta \Gamma$, where lighter shades correspond to weaker interaction. 
  Black regions represent configurations not classified as in-phase or anti-phase; they appear near the state boundary and are associated with small coupling amplitude ($\Delta \Gamma < {7\times 10^{-4}}$ 
  ).}
  \label{fig:state_map}
\end{figure}

%%%%%%%%%%%%%%%%%%%%%%

We next generalize the spatial configuration and systematically evaluate the synchronization patterns across the parameter space of the cellular orientation angles $\theta_1$ and $\theta_2$, as well as the inter-cellular distance $d$, using the phase coupling function.
By computing the phase coupling functions for all possible relative orientations, we determined the stable equilibrium phase differences for each spatial configuration. 
By symmetry under cell exchange and reflection, it is sufficient to consider
$\theta_1\in[0,\pi/2]$ and $\theta_2\in[0,\pi]$.

To construct the dynamical state maps (Fig.~\ref{fig:state_map}), we classified the stable phase differences $\Delta\phi^\ast$ obtained from phase reduction theory. Configurations were categorized as in-phase if $|\Delta\phi^\ast| < 0.1$ (blue) and anti-phase if $|\pi - |\Delta\phi^\ast|| < 0.1$ (red). The intermediate states that did not satisfy either criterion are shown in black. 
The coupling amplitude $\Delta \Gamma = \Gamma_{\max} - \Gamma_{\min}$ varies continuously and becomes small near the state boundary, where the interaction is significantly weakened and the synchronization time becomes longer.
In the intermediate states, we find $\Delta \Gamma <$ {$7\times 10^{-4}$} 
and $|\Delta\phi^\ast|$ {is often found} near $0$ and $\pi$. Thus, they represent marginal states at the state boundary rather than distinct out-of-phase states.
Comparing the results for $d/l_{\rm cell}=2.5$ and $d/l_{\rm cell}=6$, we find that the state boundary exhibits a similar shape in the $(\theta_1, \theta_2)$ plane, reflecting the robust orientation-dependence of the phase selection.

\begin{figure}[htbp]
  \centering
  \includegraphics[width=0.7\linewidth]{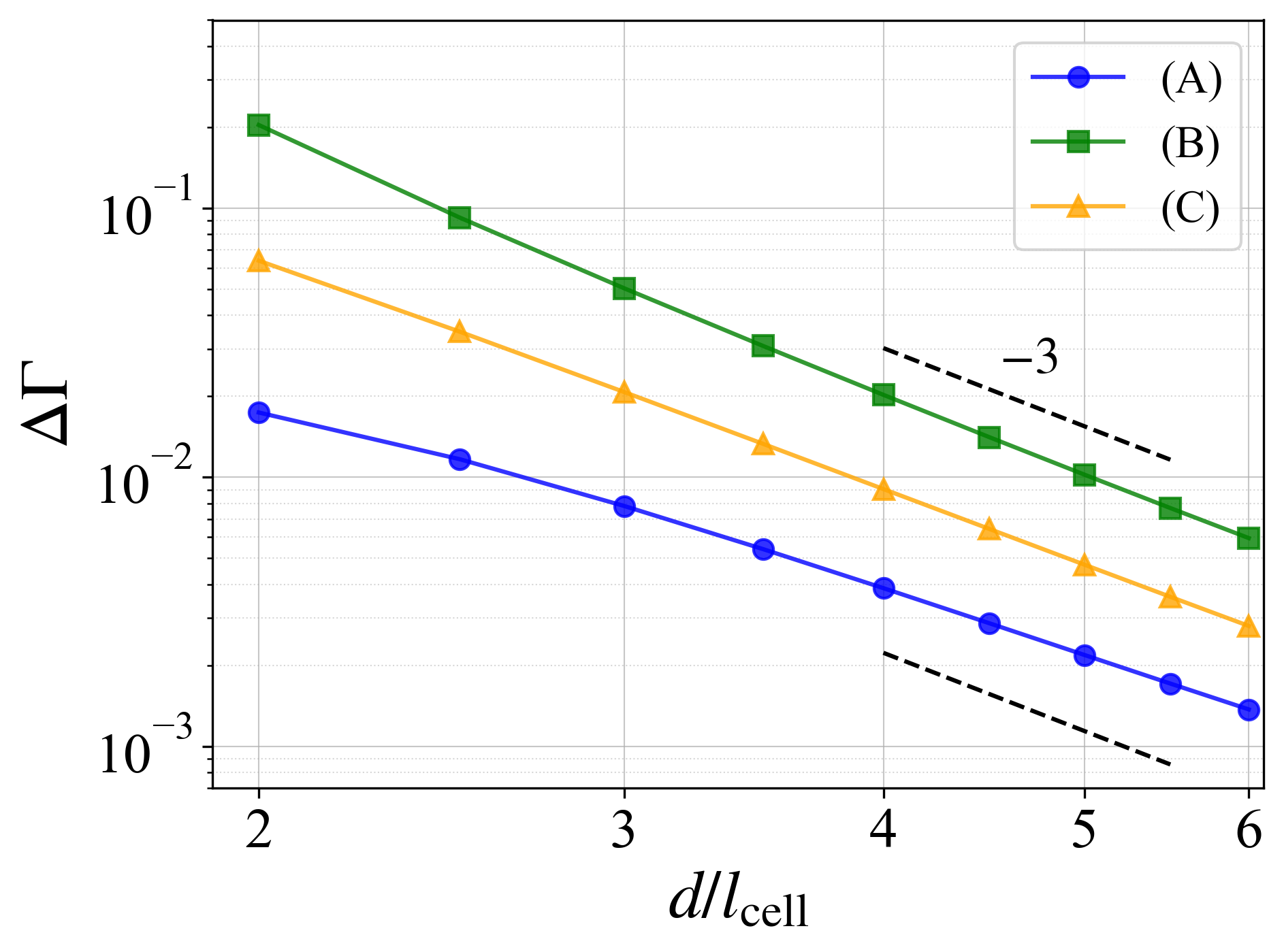}
  \caption{Log-log plot of the coupling amplitude $\Delta \Gamma$ for the \Add{same representative} configurations \Add{as in Fig.~\ref{fig:comparison}}:
  (A) $\theta_1=\theta_2=\pi/2$, (B) $\theta_1=\theta_2=0$, and (C) $\theta_1=\pi/2$, $\theta_2=0$. The dashed lines indicate the slope $-3$, corresponding to inverse cubic decay.
  }
  \label{fig:amplitude}
\end{figure}

We next examine the distance dependence of the coupling strength. As shown in Fig.~\ref{fig:amplitude}, the coupling amplitude $\Delta\Gamma$ 
decays {asymptotically} as $d^{-3}$.
This reflects the dipolar nature of the interaction and implies a long characteristic synchronization time, $\tau_{\rm sync}\sim 1/\Delta\Gamma \sim$ 
{$10$--$10^3$}.
Combined with the state maps in Fig.~3, this result suggests that the relative angular alignment primarily determines the selected synchronized state, while the spatial separation mainly affects the coupling strength and the synchronization time.

%%%%%%%%%%%%%%%%%%%%%%
%%%%%%%%%%%%%%%%%%%%%%

%%%%%%%%%%%%%%%%%%%%%%%%%%%%%%%%%%%%%%%%%%%%%%%%%%%%%%%%
%\section{Discussion}

In summary, we provide a dynamical framework for substrate-mediated synchronization of cardiomyocytes based on phase reduction theory. In particular, it predicts
the geometry-dependent selection of the phase-locked state, the coupling
strength, and the characteristic synchronization time. The orientation-dependent state map clarifies how the relative geometry of a cell pair determines the selected synchronized state, whereas the distance dependence of the coupling amplitude determines the synchronization time. The present phase-reduction framework is also reminiscent of hydrodynamic synchronization theories for cilia and flagella, where geometry-dependent coupling plays a central role
\Add{~\cite{vilfan2006,niedermayer2008,elfring2009,geyer2013,brumley2014,uchida2011,uchida2017}}.
%\Del{~\cite{uchida2011, uchida2017}}

For the representative configurations considered previously, our dynamical analysis agrees with the earlier energetic prediction {regarding the selection of in-phase or anti-phase synchronization}~\cite{cohen2016}. This agreement is nevertheless nontrivial. Indeed, in the small-deformation approximation, the substrate-mediated elastic energy can be written as $H=\frac{1}{2}\bm{X}^{\rm T}\bm{M}^{-1}\bm{X}$, so that the reaction force is conservative, $-\hat{F}_i=-\partial H/\partial X_i$. However, the synchronized state is selected through the non-isochronous part of the phase dynamics, $\dot{\phi}_i - \Omega_0 = -Z_2(\phi_i) \hat{F}_i$, where $Z_2$ denotes the second component of the phase sensitivity function. This component is not simply proportional to $\partial X/\partial \phi$ in general. Therefore, the phase velocity is not necessarily proportional to the energy gradient, $-\partial H/\partial \phi_i = \hat{F}_i (\partial X_i / \partial \phi_i)$. 
\Add{A related observation was made in hydrodynamic synchronization, where phase-locked states can correspond either to minima or maxima of the energy dissipation depending on the waveform symmetry~\cite{elfring2009}.
This example also illustrates that an energetic criterion alone does not generally determine the dynamically selected synchronized state.}
\Del{An analytical approach in simplified limits may further clarify the geometric structure of the coupling, but this lies beyond the scope of the present Letter and will be discussed elsewhere.}
\Add{Noise may obscure synchronization at large separations where $\Delta\Gamma$ 
is small. Quantitative experimental characterization of fluctuations and 
stochastic theory remain important subjects for future work.}

\smallskip

{\bf Data Availability:} The numerical data and simulation code supporting the findings of this study are available at https://doi.org/10.5281/zenodo.19547028.
%%%%%%%%%%%%%%%%%%%%%%%%%%%%%%%%%%%%%%%
%\bibliographystyle{jpsj}
%\bibliography{tomiie2026dynamical}

\end{document}